\documentclass[a4paper]{aa} 
\usepackage{graphics}
\begin{document}

\thesaurus{06
		(03.13.4;
		 05.03.1;
		 08.02.3;
		 10.07.2)}
\title{Fitting Formulae for Cross Sections of Tidal Capture Binary Formation}

\author{Sungsoo S. Kim\inst{1} \and Hyung Mok Lee\inst{2}
          }

\offprints{S.S. Kim}

\institute{Korea Advanced Institute of Science \& Technology, Department of
		Physics, Space Science Lab., Daejon 305-701, Korea\\
		email: sskim@space.kaist.ac.kr
	\and
		Seoul National University, Department of Astronomy,
		Seoul 151-742, Korea\\
		email: hmlee@astro.snu.ac.kr
             }

\date{Received March 16, 1999; accepted April 30, 1999}

\titlerunning{(RN) Cross Sections of Tidal Capture Binary}
\authorrunning{S.S. Kim \& H.M. Lee}
\maketitle

\begin{abstract}
Tidal captures can produce objects that are observationally and dynamically
important in dense stellar systems. Recent discoveries of compact young
clusters in and out of the Galaxy have prompted the studies of dynamics of
star clusters with a large range in stellar masses. The
tidal interactions between high and
low mass stars are found to be rather frequent in such clusters. 
In this Research Note, we present fitting formulae for the cross sections of 
tidal capture binary formation between two stars with a large mass ratio.
We present the cases between two main-sequence stars,
and between a degenerate star and a main-sequence star.
\keywords{methods: numerical -- celestial mechanics, stellar dynamics --
binaries: general -- globular clusters: general}
\end{abstract}

\section{INTRODUCTION}

A relatively large number of X-ray sources in globular clusters was first
pointed out by Katz (\cite{K75}).  These objects are
thought to be close binary systems. As a mechanism for the formation
of these binaries, the tidal capture of a normal star by a degenerate star
was suggested by Clark (\cite{C75}) and Fabian et al. (\cite{FPR75}).
In addition, tidally captured binaries can play an important role in
globular cluster dynamics (e.g., Ostriker \cite{O85}, Kim et al. \cite{KLG98}).

A precise mechanism for the dissipational tidal capture process was
introduced by Fabian et al. (\cite{FPR75}), and a detailed computation for the
amount of energy deposited into oscillatory modes during a close encounter
was performed by Press \& Teukolsky (\cite{PT77}) for an
$n=3$ polytropic model.  This work was extended by various
authors to other polytropes (Lee \& Ostriker
\cite{LO86}; abbreviated as LO hereafter, Ray et al. \cite{RKA87}), and
to realistic stellar models (McMillan et al. \cite{MMT87}).
Some further considerations to the subsequent dynamical evolution of tidal 
capture binaries are presented by Kochanek (\cite{K92}) and Mardling 
(\cite{M95a}, \cite{M95b}).
Many numerical simulations for close stellar encounters, whose products
include tidal capture binaries, have been performed (e.g. Benz \& Hills
\cite{BH87}; Davies et al. \cite{DBH91}).

The interests for tidal capture process have been mainly
concentrated on old stellar systems such as globular clusters and
galactic nuclei whose densities are known to be very high.
Therefore, the
cross sections were calculated only for the cases applicable to those
systems. The range of mass in these systems is considerably smaller
than that in young stellar systems.

Recent advances in infrared astronomy lead to the discoveries of
compact young clusters near the Galactic Center (Okuda et al. \cite{Oe90};
Nagata et al. \cite{Ne95}), which have been
also observed in detail using the {\it HST}
(Figer et al. \cite{Fe99}). These clusters are
found to be as dense as some globular clusters.
The sharp image quality of the {\it HST}
also enabled to find star clusters, whose age can be deduced
from the population synthesis technique, from galaxies 
at considerable distances (e.g., \"Ostlin et al. \cite{Oe98}).  

The evolution of young clusters near the Galactic center region 
becomes an important issue in 
understanding the evolutionary history of the bulge of our galaxy,
because these clusters would have rather short evaporation times.
The main reason for fast dissolution is the strong tidal field environment.
The dynamical process is further influenced by close encounters
between stars, including tidal captures.
The young clusters found in external galaxies are also of great interest
in view of the general evolution of the galaxies. Again, the dynamical
evolution and evaporation process are the key in these problems.
Two-body relaxation drives the dynamical evolution and close interactions
between two stars modify the course of evolution significantly
(see, for example, Meylan \& Heggie \cite{MH97}).


The effect of tidal interaction during stellar encounter can be incorporated
in different ways depending on the methods of the study of
dynamical evolution of stellar systems. In direct N-body calculations,
one should know exactly when the tidal capture takes place. In statistical
methods such as Fokker-Planck models, the capture cross section as a function
of other kinematic parameters (usually velocity dispersion) is necessary.

In the present Research Note, we present convenient formulae for
cross sections for tidal capture between two stars
as a function of relative velocity at infinity.
These formulae can be easily incorporated into
statistical methods such as Fokker-Planck models or gaseous models
for the studies of dynamics of star clusters. They will also be useful
in making estimates for the interaction rates for a given cluster
parameters.

\section{CROSS SECTIONS}

Previous studies presented tidal cross sections for the limited
ranges of stars. For example, the mass ratios between encountering stars 
considered in LO ranged only
from 1 to 8 for normal-normal star pairs, and 0.5 to 1.5 for
normal-degenerate pairs.
Also, $R_1/M_1=R_2/M_2$ was assumed and only encounters between the
same polytropes were considered in LO, which may not be
appropriate for encounters between stars with large mass ratios.
While an $n=1.5$ polytrope may well represent the
structure of low-mass stars, the outer structure of intermediate to massive
stars ($M {\mathrel{\hbox to 0pt{\lower 3.0pt\hbox{$\mathchar"218$}\hss}
\raise 2.0pt\hbox{$\mathchar"13E$}}} 1 \, M_\odot$)
may be better represented by a $n=3$ polytrope.  Thus a consideration
for the encounter between $n=1.5$ and 3 polytropes is necessary for
encounters with a large mass difference.
Here we extend the work of LO to obtain the cross sections for
i) encounters between stars with a very large mass ratio,
ii) encounters with mass-radius relation that deviates from 
conventional $R\propto M$ relation (i.e., 
$[R_2/M_2]/[R_1/M_1]$ values other than 1), and
iii) encounters between $n=1.5$ and 3 polytropes, and
present the results in the form of convenient
fitting formulae for cross sections and critical periastron distances.

The amount of orbital energy deposited to the stellar envelope
is a very steep function of the distance between stars.  The relative
orbit can be described as a function of energy and angular momentum. 
However, the relative orbit near the periastron passage 
can be approximated by a parabolic orbit which
can be specified by only one parameter: periastron distance $R_{\rm min}$.
There exists a critical $R_{\rm min}$ below which the tidal interaction
transforms the initial unbound system into a bound one for a given set
of mass and radius pair, and the relative velocity at infinity, $v_\infty$.

Assuming a parabolic relative orbit for the encounter,
Press \& Teukolsky (\cite{PT77}) expressed the deposition of kinetic energy
into stellar oscillations of a star with $M_1$ and $R_1$ due to a perturbing
star with $M_2$ and $R_2$ by
\begin{equation}
\label{deltaE}
	\Delta E_1 = \left ( {GM_1 \over R_1} \right )^2 \left ( {M_2 \over
			M_1} \right )^2 \sum_{l=2}^{\infty} \left (
			{R_1 \over R_{\rm min}} \right )^{2l+2} T_l(\eta_1),
\end{equation}
where $l$ is the spherical harmonic index, $R_{\rm min}$ the apocenter
radius, and the contribution of the
summation behind $l=3$ to $\Delta E$ is negligible. The dimensionless
parameter $\eta$ is defined by
\begin{equation}
\label{eta}
	\eta_1 \equiv \left ( {M_1 \over M_1+M_2} \right )^{1/2} \left (
			{R_{\rm min} \over R_1} \right )^{1.5}.
\end{equation}
Expression for $\Delta E_2$ may be obtained by exchanging subscripts 1 and 2
in Eqs.~(\ref{deltaE}) and (\ref{eta}).  For $T_2(\eta)$ and $T_3(\eta)$
values, we use Portegies Zwart \& Meinen's (\cite{PZM93}) fifth order fitting
polynomials to the numerical calculations by LO.

Tidal capture takes place when the deposition of kinetic energy during the
encounter 
$\Delta E = \Delta E_1 + \Delta E_2$
is larger than the initially positive orbital energy
$E_{\rm orb} = {1 \over 2} \mu  v_{\infty}^2$,
where $\mu$ is the reduced mass of $M_1$ and $M_2$ pair.
The critical $R_{\rm min}$ below which the tidal energy exceeds the
orbital energy depends on $v_\infty$ via $\eta$.

After obtaining critical $R_{\rm min}$ by requiring 
$\Delta E = {1 \over 2} \mu v_\infty^2$, we can compute the critical
impact parameter $R_0$ that leads to the critical $R_{\rm min}$ (assuming the
orbit does not change in the presence of tidal interaction), as a
function of $R_{\rm min}$ and $v_\infty$:
\begin{equation}
\label{R0}
R_0(v_{\infty}) = R_{\rm min} \left (1+{2GM_{\rm T} \over R_{\rm min}
			v_{\infty}^2} \right )^{1/2},
\end{equation}
where $M_T=M_1+M_2$. The capture cross section, $\sigma(v_{\infty})$, is simply
the area of the target, $\pi R_0^2$.
The velocity dependence of cross section
is expected to be close to $v_\infty^{-2}$ since the second term in 
the parenthesis of Eq. (\ref{R0}) is much greater than 1, and $R_{\rm min}$ is
only weakly dependent on $v_\infty$.

%
%

\section{FITTING FORMULAE}
\label{sec:formulae}

The cross section depends on $v_\infty$ in a somewhat complex way.
However, for the limited range of $v_{\infty}$, it can be approximated
as a power law on the tidal capture cross section.
Following LO, we provide $\sigma$
as a function of $R_1$ and escape velocity at the surface of star~1,
$v_{*1} \equiv (2GM_1/R_1)^{1/2}$:
\begin{equation}
	\sigma = a \left ( {v_{\infty} \over v_{*1}} \right )^{-\beta} R_1^2.
\end{equation}
We obtain constants $a$ and $\beta$ by fitting the power law curve to
$\sigma(v_{\infty})$ at $v_{\infty}=10\,{\rm km \, s^{-1}}$, which is
typical velocity range in the globular clusters and compact young clusters.
The galactic nuclei are also thought to be dense enough for tidal
interactions, but the direct collision is more probable because of
high velocity dispersion (Lee \& Nelson \cite{LN88}). 

We assumed $R_1/M_1= R_\odot / M_\odot$ for the calculation of $v_{*1}$,
but $a$ and $\beta$ values are nearly insensitive to the
choice of $R_1/M_1$ value because the above power law holds
for a wide range of $v_{\infty}$ near $10\,{\rm km \, s^{-1}}$.

\begin{figure}
\resizebox{8cm}{!}{\includegraphics{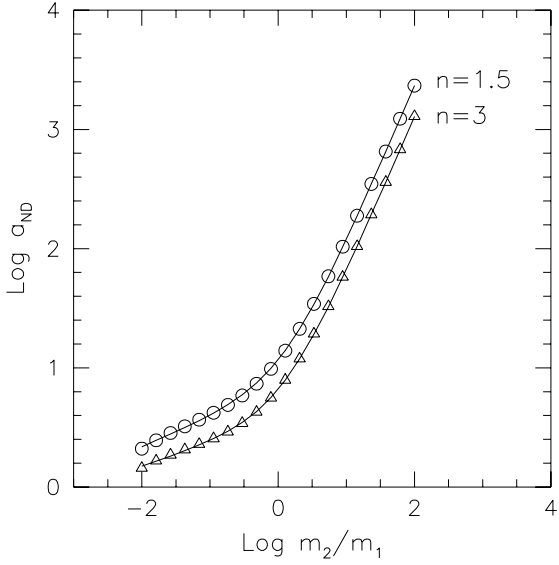}}
\caption{Calculated $a_{\rm ND}$ for $n=1.5$ (circles) and $n=3$ (triangles)
as a function of $m_2/m_1$.  Lines represent the fitting formulae
of Eq. \ref{nd}.}
\label{fig:nd}
\end{figure}

\subsection{Normal-Degenerate Encounters}
For encounters between a normal and a degenerate star, we obtain $a$ and
$\beta$ for $0.01 \leq M_2/M_1 \leq 100$.
In this subsection, subscript 1 is for the
normal star and 2 for the degenerate star.  While $a$ is a steep function of
$M_2/M_1$ (see Fig.~\ref{fig:nd}), $\beta$ ranges only from 2.24 ($M_2/M_1
= 0.01$) to 2.13 ($M_2/M_1=100$) for $n=1.5$, and from 2.24 to 2.19 for $n=3$.
Thus $\beta = 2.2$ would be a good choice.  We find that
$a$ is well fit with a sum of two power law curves:
\begin{eqnarray}
\label{nd}
	a_{\rm ND} & = 6.60 \left ({M_2\ \over M_1} \right )^{0.242} +
		       5.06 \left ({M_2\ \over M_1} \right )^{1.33}
		   & \; \; \; {\rm for} \; n=1.5; \nonumber\\
	a_{\rm ND} & = 3.66 \left ({M_2\ \over M_1} \right )^{0.200} +
		       2.94 \left ({M_2\ \over M_1} \right )^{1.32}
		   & \; \; \; {\rm for} \; n=3,
\end{eqnarray}
where subscript ND is for normal-degenerate star encounters.  Eq.~(\ref{nd})
fits our calculated $a_{\rm ND}$ values with a relative error
($|$fit-data$|$/data) better than 4~\%.

\begin{figure}
\resizebox{7.5cm}{!}{\includegraphics{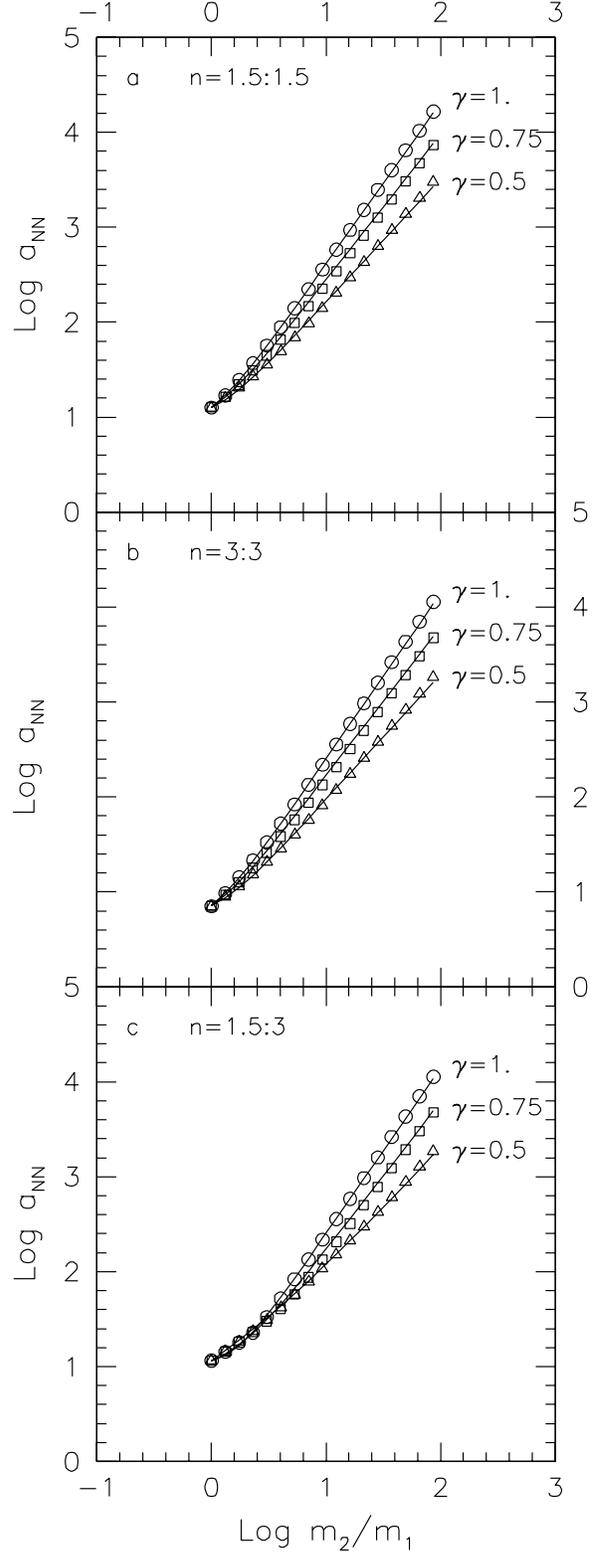}}
\caption{Calculated $a_{\rm NN}$ for $n=1.5:1.5$ ({\bf a}), $n=3:3$ ({\bf b}),
and $n=1.5:3$ ({\bf c}) as a function of $m_2/m_1$.  Circles are for
$\gamma=1$, squares for $\gamma=0.75$, and triangles for $\gamma=0.5$.
Lines represent the fitting formulae of Eq. \ref{nn}.}
\label{fig:nn}
\end{figure}

\subsection{Normal-Normal Encounters}
For encounters between normal stars, we consider $M_2/M_1$ of 1 through 100.
In this subsection, subscript 1 is for less massive star.  Encounters
between normal stars involve two more parameters, $R_2$ and $M_2$, but
only one parameter, $\gamma \equiv \log (R_2/R_1) / \log (M_2/M_1)$,
is enough in adding the second normal star.
For $0.5 \leq \gamma \leq 1$, it is found that $\beta$
ranges from 2.12 ($M_2/M_1=1$) to 2.24 ($M_2/M_1=100$) for $n=1.5:1.5$
and $n=1.5:3$, and from 2.18 to 2.24 for $n=3$.
Again, $\beta=2.2$ would be a good approximation.
Fig.~\ref{fig:nn} shows $a$ for three different
$\gamma$ values.  We find that a simple modification to the form
of Eq.~\ref{nd} can fit $a$ as a function of both $M_2/M_1$ and
$\gamma$:
\begin{eqnarray}
\label{nn}
	a_{\rm NN} = & 6.05 \left ({M_2\ \over M_1}
		                    \right )^{0.835 \, \ln \gamma+0.468} &
			\nonumber\\
		     & + \; 6.50 \left ({M_2\ \over M_1}
		                    \right )^{0.563 \, \ln \gamma+1.75}  
		       & {\rm for} \; n=1.5:1.5; \nonumber\\
	a_{\rm NN} = & 3.50 \left ({M_2\ \over M_1}
		                    \right )^{0.814 \, \ln \gamma+0.551} &
			\nonumber\\
		     & + \; 3.53 \left ({M_2\ \over M_1}
		                    \right )^{0.598 \, \ln \gamma+1.80}  
		       & {\rm for} \; n=3:3; \nonumber\\
	a_{\rm NN} = & 7.98 \left ({M_2\ \over M_1}
		                    \right )^{-1.23 \, \ln \gamma-0.232} &
			\nonumber\\
		     & + \; 3.57 \left ({M_2\ \over M_1}
		                    \right )^{0.625 \, \ln \gamma+1.81}  
		       & {\rm for} \; n=1.5:3, 
\end{eqnarray}
where subscript NN is for normal-normal star encounters, and star~1 has
$n=1.5$ in case of $n$=1.5:3 encounters.
Eq.~(\ref{nn}) fits our calculated $a_{\rm NN}$ values
with a relative error better than 10~\%.

\section{DISCUSSION}

We expressed $\sigma$ in terms of $v_{*1}$ and $R_1$ following LO
in Sect. \ref{sec:formulae}, but when $\sigma_{\rm NN}$ is expressed
in terms of $v_{*2}$ and $R_2$ such that
\begin{equation}
	\sigma_{\rm NN} = a'_{\rm NN} \left ( {v_{\infty} \over v_{*2}}
				      \right )^{-\beta} R_2^2,
\end{equation}
the $\gamma$ dependence of $a'_{\rm NN}$ becomes smaller than that of
$a_{\rm NN}$.
Also note that $a'_{\rm NN}(M_2/M_1)$ for $\gamma =1$ is nearly the same as
$a_{\rm ND}(M_1/M_2)$ with only a slight difference near $M_2/M_1 \simeq 1$.

Critical $R_{\rm min}$ for tidal captures is also frequently useful for some
studies such as the ones with N-body methods.  One finds the approximate
critical $R_{\rm min}$ as
\begin{equation}
	R_{\rm min} \simeq {a \over \pi} \left ( {R_1 \over 1+M_2/M_1} \right )
		  \left ( {v_{\infty} \over v_{*1}} \right )^{2-\beta}
\end{equation}
for the velocities that satisfy
\begin{equation}
	\left ( {v_\infty \over v_{*1}} \right )^{\beta -4} \gg
		{4a \over \pi} \left ( {1 \over 1+M_2/M_1} \right )^2,
\end{equation}
where the right-hand-side does not vary much from 10.

Heating by tidal capture binaries is incorporated in Fokker-Planck models
by calculating $\langle \sigma v \rangle$, where brackets indicate the average
over velocity.  With the Maxwellian velocity distribution for stars,
we obtain $\langle \sigma v \rangle \propto \langle v^{-1.2} \rangle
\simeq 1.5 v_{\rm rms}^{-1.2}$ where $v_{\rm rms}$ is the root-mean-square
relative velocity between two stellar mass groups.
The $\sigma$ presented here also includes encounters that will
lead to a merge before encountering a third star.  We will not, however,
attempt to go over this issue because it is beyond our scope in this
Research Note.  We find that the maximum velocity beyond which $R_0 \leq R_1$
is significantly larger than 100~${\rm km \, s^{-1}}$ for our parameter regime.

In this Research Note, we merely gave the cross sections. The subsequent
evolution of tidally captured systems is a very important subject, but
is a rather difficult to follow. If the energy deposited to the
envelope of stars can be quickly radiated away before two stars
become close, the final orbit of the binary will be a circle whose
radius is twice of the initial $R_{\rm min}$ (LO). The stellar rotation
induced by the tidal interactions during the circularization process
can reduce the separation of circularized orbits. Therefore, the tidal
products are usually stellar mergers or tight binaries.

However, there are possibilities of resonant interactions between
the tides and stellar orbits. In such an environment, the tidal 
energy can also be transferred back to the orbital energy. The
final product of such a case is a rather wide binary, although it
is still ``hard" binary in terms of cluster dynamics (Kochanek \cite{K92},
Mardling \cite{M95a}, \cite{M95b}).
Therefore, the tidal capture could produce dynamically
and observationally interesting objects in dense stellar environments.
The cross sections presented here can be
useful in estimating the frequencies of such interactions.

\begin{acknowledgements}
This work was supported in part by the International Cooperative Research
Project of Korea Research Foundation to Pusan National University, in 1997.
\end{acknowledgements}


\end{document}